\documentclass[epsf]{aa}
\usepackage{natbib}
\usepackage{epsfig}

\newcommand{\teff}{\mbox{$T_{\rm eff}$}}

\newcommand{\kms}{\mbox{km\,s$^{-1}$}}

\def\ms{\,m\,s$^{-1}$}         %m.s -1
\def\kms{\,km\,s$^{-1}$}       %km.s -1
\def\teff{$T_{\rm eff}$}

\def\kms{km\, s$^{-1}$}

\begin{document}
%IRAC%%%%%%%%%%%%%%%%%%%%%%%%%%%%%%%%%%%%%%%%%%%%%%%%%%%
\title{The thermal emission of the young and massive  planet CoRoT-2b at 4.5 and 8 $\mu$m\thanks{Based on data collected with the VLT/FORS2 instrument at ESO Paranal Observatory, Chile (programs 081.C-0413(B)).}$^{, }$\thanks{The photometric time-series used in this work are only available in electronic form at the CDS via anonymous ftp to  cdsarc.u-strasbg.fr (130.79.128.5) or via http://cdsweb.u-strasbg.fr/cgi-bin/qcat?J/A+A/}}
\author{M.~Gillon$^{1,2}$,  A.~A.~Lanotte$^{1}$, T.~Barman$^{3}$, N.~Miller$^{4}$, B.-O.~Demory$^2$, M.~Deleuil$^5$, J.~Montalb\'an$^{1}$,  F.~Bouchy$^6$, A.~Collier~Cameron$^7$,  H.~J.~Deeg$^8$, J.~J.~Fortney$^4$, M.~Fridlund$^9$, J.~Harrington$^{10}$, P.~Magain$^1$, C.~Moutou$^5$, D.~Queloz$^2$, H.~Rauer$^{11,12}$, D.~Rouan$^{13}$, J.~Schneider$^{14}$}   

\offprints{michael.gillon@ulg.ac.be}
\institute{
$^1$ Institut d'Astrophysique et de G\'eophysique,  Universit\'e
  de Li\`ege,  All\'ee du 6 Ao\^ut 17,  Bat.  B5C, 4000 Li\`ege, Belgium \\
$^2$  Observatoire de Gen\`eve, Universit\'e de Gen\`eve, 51 Chemin des Maillettes, 1290 Sauverny, Switzerland\\
$^3$ Lowell Observatory, 1400 West Mars Hill Road, Flagstaff, AZ 86001, USA\\
$^4$ Department of Astronomy and Astrophysics, University of California, Santa Cruz, USA\\
$^5$ LAM, UMR 6110 CNRS, 38  rue Fr\'ed\'eric Joliot-Curie, 13388 Marseille, France\\
$^6$ Observatoire de Haute Provence, USR 2207 CNRS, OAMP, F-04870 St-Michel l'Observatoire, France\\
$^7$ School of Physics and Astronomy, University of St. Andrews, North Haugh, Fife, KY16 9SS, UK\\
$^8$ Instituto de Astrof\'isica de Canarias, C. Via Lactea S/N, E-38200 La Laguna, Spain\\
$^9$ Research and Scientific Support Department, European Space Agency, ESTEC, 220 Noordwijk, The Netherlands\\
$^{10}$ Planetary Sciences Group, Department of Physics, University of Central Florida, Orlando, FL 32816, USA\\
$^{11}$ Institut fuer Planetenforschung, DLR, Rutherford str. 2,12489 Berlin, Germany\\ 
$^{12}$ Zentrum fuer Astronomie und Astrophysik, Hardenbergstr. 36,10623 Berlin, Germany\\ 
$^{13}$ LESIA, UMR 8109 CNRS, Observatoire de Paris, UVSQ, Universit\'e Paris-Diderot, 5 Place J. Janssen, 92195 Meudon, France\\
$^{14}$ LUTH, UMR 8102 CNRS, Observatoire de Paris-Meudon, 5 Place J. Janssen, 92195 Meudon, France\\
}

\date{Received date / accepted date}
\authorrunning{M. Gillon et al.}
\titlerunning{The thermal emission of CoRoT-2b at 4.5 and 8 $\mu$m}
%%%%%%%%%%%%%%%%%%%%%%%%%%%%%%%%%%%%%%%%%%%%%%%%%%%%
\abstract{ We report measurements of the thermal emission of the young and massive planet CoRoT-2b at 4.5 and 8 $\mu$m with the  $Spitzer$ Infrared Array Camera (IRAC). Our measured occultation depths are $0.510\pm0.042$ \% and $0.41\pm0.11$ \% 
at 4.5 and 8 $\mu$m, respectively. In addition to the CoRoT optical measurements, these planet/star flux ratios indicate 
a poor heat distribution to the night side of the planet and are in better agreement with an atmosphere free of temperature inversion layer. Still, the presence of such an inversion is not definitely ruled out by the observations and a larger wavelength coverage is required to remove the current ambiguity. Our global analysis of CoRoT, $Spitzer$ and ground-based data confirms the large mass and size of the planet with slightly revised values ($M_p=3.47\pm0.22$~$M_J$, $R_p=1.466\pm0.044$~$R_J$).  We find a small but significant offset in the timing of the occultation when compared to a purely circular orbital solution, leading to $e \cos{\omega} =-0.00291\pm0.00063$ where $e$ is the orbital eccentricity and  $\omega$ is the argument of periastron. Constraining the age of the system to be at most of a few hundreds of Myr and assuming that the non-zero orbital eccentricity is not due to a third undetected body, we model the coupled orbital-tidal evolution of the  system with various tidal $Q$ values, core sizes  and initial orbital parameters. For $Q_s' = 10^5 - 10^6$, our modelling is able to explain the large radius of CoRoT-2b if $Q_p' \le 10^{5.5}$ through a transient tidal circularization and corresponding planet tidal heating event. Under this model, the planet will reach its Roche limit within 20 Myr at most.

  }
%%%%%%%%%%%%%%%%%%%%%%%%%%%%%%%%%%%%%%%%%%%%%%%%%%%%

\maketitle

\section{Introduction}

Transiting planets are key objects for our understanding of the atmospheric properties of exoplanets. Indeed, their special geometrical configuration gives us the opportunity not only to deduce their density but also to study directly their atmospheres without the challenging need to spatially resolve their light from that of their host star. In particular, their emergent flux can be directly measured during their occultation (secondary eclipse) when they are hidden by their host star, as was demonstrated by Charbonneau et al. (2005) and  Deming et al. (2005). Since 2005, many exoplanet occultation  measurements have been gathered, the bulk of them by  the $Spitzer$ $Space$ $Telescope$ (see, e.g., Deming 2009), the few others being due to the $Hubble$ $Space$ $Telescope$ (Swain et al. 2009), CoRoT (Alonso et al. 2009b, 2009c; Snellen et al. 2009a), Kepler (Borucki et al. 2009) and  ground-based telescopes (Sing \& L\'opez-Morales 2009, de Mooij \& Snellen 2009, Gillon et al. 2009b). Combining the photometric measurements at different wavelengths allows us to map the  Spectral Energy Distribution (SED) of the planet and to constrain its chemical composition, its thermal distribution efficiency and the presence of a possible stratospheric thermal inversion (see, e.g., Charbonneau et al. 2008, Knutson et al. 2008). Such inversions have been detected for the highly irradiated planets HD 209458b (Burrows et al. 2007b, Knutson et al. 2008), TrES-2b (O'Donovan et al. 2009), TrES-4b (Knutson et al. 2009), XO-1b (Machalek et al 2008) and XO-2b (Machalek et al. 2009). These results are in rather good agreement with the theoretical division of hot Jupiters into two classes based on their level of irradiation (Hubeny et al. 2003, Burrows et al. 2007b, Harrington et al. 2007, Burrows et al. 2008, Fortney et al. 2008). Under this division, the planets warmer than required for condensation of high-opacity gaseous molecules like TiO/VO, tholins or polyacetylenes should show a stratospheric temperature inversion due to the absorption in their upper-atmosphere of a significant fraction of the large incident flux by these  compounds. The less irradiated planets would lack these gazeous compounds and the resulting temperature inversion. Still, this simple division was recently challenged by the absence of thermal inversion reported for the strongly irradiated planet TrES-3b by Fressin et al. (2009). This result indicates that, in addition to the  irradiation amplitude, other effects like chemical composition, surface gravity and the stellar spectrum have probably an impact on the temperature profile of hot Jupiters.

 With an irradiation $\sim1.3\times10^9$ erg~s$^{-1}$~cm$^{-2}$, the planet CoRoT-2b could be expected to show such a temperature inversion, according to the theoretical division mentionned above. This planet is the second one discovered by the CoRoT transit survey mission (Alonso et al. 2008, hereafter A08). Spectral analysis and evolution modelling of the host star leads to a solar-type dwarf with a mass $M_\ast=0.97\pm0.06$~$M_\odot$ and an effective temperature \teff$=5625\pm120$~K (Bouchy et al. 2008, hereafter B08).  A08 derived for the planet a radius of $1.465\pm0.029$~$R_{J}$ and a mass of $3.31\pm0.16$~$M_{J}$, leading to a density of $1.31\pm0.04$~g cm$^{-3}$, very close to the value for Jupiter. This density is surprising because the radius of massive planets is expected to approach Jupiter's asymptotically. In this context, it is worth noticing the probably young age of the system. Indeed, the presence in the stellar spectrum of the Li I absorption line and the strong emission in the Ca II H and K line cores (B08) suggest that  the star is still close to the Zero-Age Main-Sequence (ZAMS) and is thus younger than 0.5 Gyr (B08), in full agreement with the short rotational period of $\sim4.5$ days deduced from CoRoT photometry (A08). Still, the youngness of CoRoT-2b is not enough to prevent it from falling into the sub-group of planets with a radius larger than predicted by basic models of irradiated planets (Burrows et al. 2007a, Fortney et al. 2007). Most of these planets show an orbital eccentricity compatible with zero. Nevertheless, these planets could still have undergone during their evolution a tidal heating large enough to explain their low density (Jackson et al. 2008b; Ibgui \& Burrows 2009), and it is thus important to measure very precisely their present eccentricity to constrain their tidal and thermal history. The precise measurement of a planet's occultation provides strong constraints on the orbital eccentricity, especially on the parameter $e \cos{\omega}$, where $e$ is the eccentricity and $\omega$ the argument of periastron (see e.g. Charbonneau et al. 2005, Knutson et al. 2009). In the case of CoRoT-2b, the dynamical interest of such occultation observations is reinforced by the large jitter noise of the young host star (B08), which makes the precise determination of a tiny eccentricity very challenging for the radial velocity (RV) method alone. 

 With the goals of better characterizing the atmospheric properties of CoRoT-2b (SED, inversion) and improve our understanding of its low density (tidal heating), we observed the occultation of this planet at 4.5 and 8 $\mu$m with $Spitzer$/IRAC (DDT program 486). A partial transit was also observed with VLT/FORS2 to bring one more constraint on the orbital parameters.  We report here the results of the analysis of these new data. Section 2 presents our IRAC and VLT observations and their reduction. We analyzed this new photometry in combination with CoRoT transit photometry and published RVs. This combined analysis is presented in Sec. 3. We present and discuss our results in Sec. 4 and give our conclusions in Sec. 5.

\section{New photometric observations}

\subsection{IRAC occultation photometry}

CoRoT-2 (2MASS 19270649+0123013, $K_s$ = 10.31) was observed by $Spitzer$ (Werner et al. 2004) during an occultation of its planet on November $1^{st}$ 2008 from 03h50 to 08h50 UT. The observations were performed with the Infrared Array Camera (IRAC) (Fazio et al. 2004) in full array mode ($256\times256$ pixels, 1.2 arcsec/pixel) simultaneously at 4.5 and 8 $\mu$m. The telescope was not repointed during the course of the observations to minimize the motion of the stars on the array. We carefully selected the pointing in order (1) to avoid the bright star 2MASS 19270954+0123280 ($K_s$ = 7.55) that would have saturated the detector for any exposure time while ensuring that it will not fall into one of several regions outside the FOV that are known to result in significant scattered light on the detectors, and (2) to avoid areas of the array with known bad pixels or significant gradients in the flat field as well as areas known to be affected by scattered starlight. We ensured also that no bright star would have been located into stray light avoidance zones\footnote{For details, see the IRAC Data Handbook available at http://ssc.spitzer.caltech.edu/irac}. An effective integration time of 10.4s was used during the whole run, resulting in 1385 images for each channel. For our analysis, we used the images calibrated by the standard $Spitzer$ pipeline (version S18.0) and delivered to the community as Basic Calibrated Data (BCD). We converted fluxes from the $Spitzer$ units of specific intensity (MJy/sr) to photon counts, and aperture photometry was obtained for CoRoT-2 in each image using the {\tt IRAF/DAOPHOT}\footnote{{\tt IRAF} is distributed by the National Optical Astronomy Observatory, which is operated by the Association of Universities for Research in Astronomy, Inc., under cooperative agreement with the National Science Foundation.} software (Stetson, 1987). In both channels, the Point-Spread Function (PSF) of the target is slightly blended with the one of the fainter ($K_s$ = 12.03) redder ($J - K_s$ = 0.84 $vs$ 0.47 for CoRoT-2) star 2MASS 19270636+0122577 located at $\sim$ 4'' (see Fig. 1). A small aperture radius was used for both channels (4.5 $\mu$m: 4 pixels, 8 $\mu$m: 3.5 pixels). The aperture was centered in each image by fitting a Gaussian profile on CoRoT-2. A mean sky background was measured in an annulus extending from 8 to 16 pixels from the center of the aperture, and subtracted from the measured flux for each image. Each measurement was compared to the median of the ten adjacent images and rejected as an outlier if the difference was larger than four times its theoretical error bar. Twenty-four points (3.5 \%) were rejected at 4.5 $\mu$m and 37 points (2.7 \%) at 8 $\mu$m. Figure 2 shows the resulting time-series for both channels. Despite its small size, the photometric aperture does not only contain counts due to CoRoT-2 but also to the nearby fainter star, leading to a dilution of the eclipse. To estimate this dilution and correct the measured eclipse depths for it, we used the following procedure. For both channels, we partially deconvolved the images taken after the occultation, using the deconvolution program {\tt DECPHOT} (Gillon et al. 2006, 2007a, Magain et al. 2007). We used the oversampled high-SNR PSF available on $Spitzer$'s web site\footnote{http://ssc.spitzer.caltech.edu/irac/psf.html} to deduce the partial PSF needed for this deconvolution. The deconvolved images (see Fig. 1) are oversampled by a factor of 2 and their PSF is a Gaussian with a Full-Width at Half Maximum (FWHM) of 2 pixels, corresponding thus to a FWHM of 1 pixel for the original sampling. This has to be compared to a FWHM $\sim$ 1.5 pixel for the original images. We performed aperture photometry on the PSF model to measure the fraction $\alpha$ of the flux of CoRoT-2  within an aperture of 8 (4.5 $\mu$m) and 7 (8 $\mu$m) pixels. At this stage, we compared the total flux $F_{tot}$ of CoRoT-2 obtained by {\tt DECPHOT} for each image to the flux obtained with aperture photometry on the model images (i.e. the obtained higher-resolution images convolved by the partial PSF model). This last measurement should be the sum of $\alpha \times F_{tot}$ and the contaminating flux due to the nearby star. Subtracting $\alpha \times F_{tot}$ to this quantity and dividing by the same  $\alpha \times F_{tot}$ finally gave us an estimation of the aperture contamination due to the nearby star. Considering all the images taken after the occultation, we obtained a dilution of $16.4\pm0.4$ \% and $14.3\pm0.7$ \%, respectively at 4.5 and 8 $\mu$m. 

\begin{figure}
\label{fig:a}
\centering                     
\includegraphics[width=8cm]{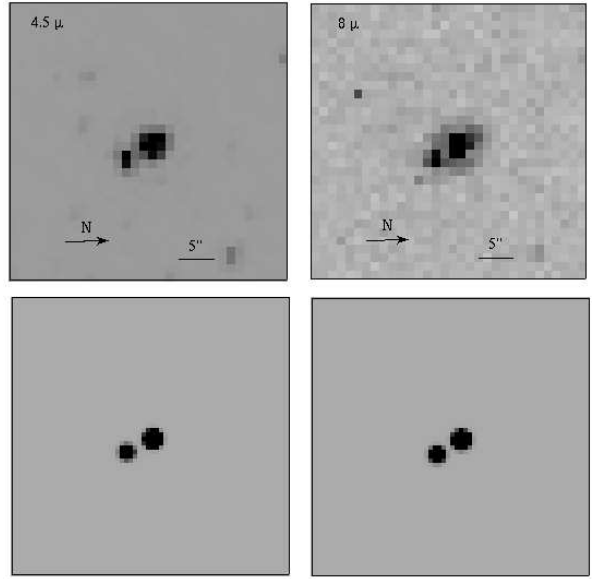}
\caption{$Top$: Zoom on CoRoT-2 and the nearby fainter star within an IRAC image taken at 4.5 $\mu$m ($left$) and 8 $\mu$m (right). $Bottom$: Same, but after deconvolution. }
\end{figure}

\begin{figure}
\label{fig:b}
\centering                     
\includegraphics[width=9cm]{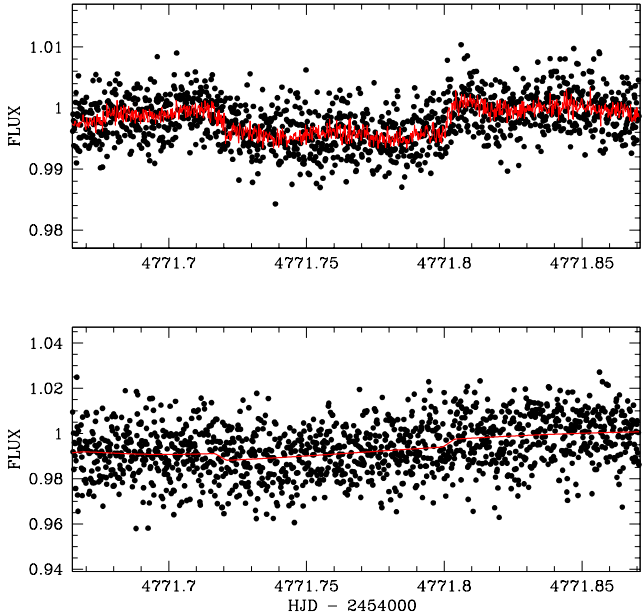}
\caption{IRAC occultation photometry obtained at 4.5 $\mu$m ($top$) and 8 $\mu$m ($bottom)$. For both time-series, the best-fitting occultation+systematics model is superimposed (in red).  }
\end{figure}

\subsection{VLT/FORS2 transit photometry}

A partial transit of CoRoT-2b was observed on September 9$^{th}$ 2008 with the FORS2 camera (Appenzeller et al. 1998) installed at the VLT/UT1 (Antu). The FORS2 camera has a mosaic of two 2k $\times$ 4k MIT CCDs and is optimized for observations in the red with a very low level of fringes. It was used several times in the past to obtain high precision transit photometry (e.g. Gillon et al. 2007b, 2009a, 2009b; Pont et al. 2007). The high resolution mode and 1$\times$1 binning were used to optimize the spatial sampling, resulting in a 4.6' $\times$ 4.6' field of view with a  pixel scale of  0.063''/pixel. Airmass decreased from 1.18 to 1.11 then increased to 1.35 during the run which lasted  from 23h40 to 3h12 UT. The quality of the night was photometric. 

We acquired 448 images in the  $z$-GUNN+78 filter ($\lambda_{eff}= 910 $ nm, FWHM = 130.5 nm) with an exposure time ranging from 0.6 to 3~s. After a standard pre-reduction,  the stellar fluxes were extracted for all the images with the {\tt IRAF/DAOPHOT}  aperture photometry software. Fifty images were revealed to be saturated and were discarded from the analysis. Several sets of reduction parameters were tested, and we kept the one giving the most precise photometry for the stars of similar brightness than CoRoT-2. After a careful selection of reference stars, differential photometry was obtained.  The resulting transit light curve is shown in Fig. 3. After subtraction of the best-fit model (see next section), the obtained residuals show a standard deviation of $\sim 1.9 \times 10^{-3}$,  close to the mean theoretical noise ($\sim 1.7 \times 10^{-3}$).

\begin{figure}
\label{fig:c}
\centering                     
\includegraphics[width=9cm]{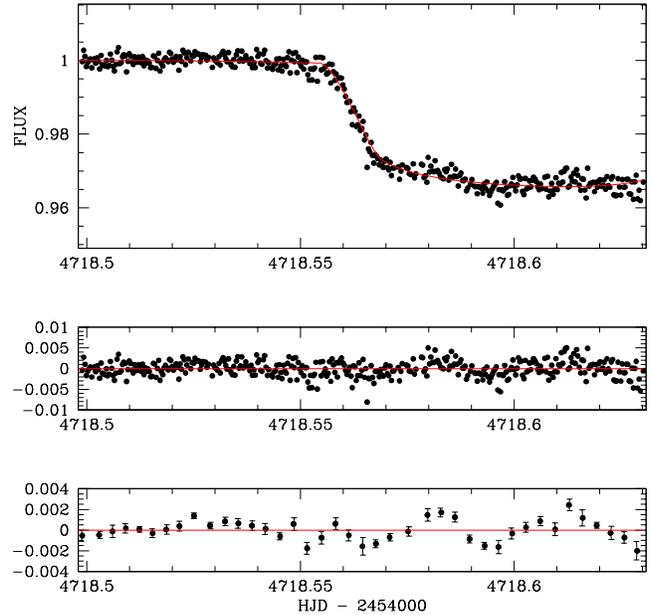}
\caption{$Top$: VLT/FORS2 $z$-band transit photometry with the best-fitting transit+systematics model superimposed (in red). $Middle$ and $bottom$: Residuals of the fit unbinned and binned per 20 minutes. The larger scatter during the transit is probably due to the inhomogeneity of the stellar surface (spots).}
\end{figure}

\section{Data analysis}

\subsection{Used data and model}

We performed a global determination of the system parameters based on our new photometry in addition to the following data: 
\begin{itemize}
\item The phase-folded CoRoT transit photometry presented in A08. The 160 measurements of this transit light curve were obtained after folding the 78
 transits observed by CoRoT using the precise ephemeris deduced in A08 and after binning the resulting light curve with a bin size of $\sim$ 2.5 min. 
 For our analysis, we projected this phase-folded photometry onto the central transit timing presented in A08. To take into account the 
 uncertainty on the time of minimum light  and  on the orbital period presented in A08 ($T_0=2454237.53562\pm0.00014$ HJD and 
 $P=1.7429964\pm0.0000017$ days), new values were randomly drawn from the corresponding normal distributions at the beginning of 
 each chain of the MCMC analysis (see below) before projecting the phase-folded light curve. 
 \item The radial velocity (RV) measurements published in A08 and B08 and obtained by the HARPS and SOPHIE spectrographs. These RVs encompass two 
transits. These spectroscopic transit observations were obtained to measure the sky-projected angle $\beta$ between 
the planetary orbital axis and the stellar rotation axis via the observation of the Rossiter-McLaughlin effect (RM; Queloz et al. 2000).
We included these spectroscopic transit observations in our analysis to benefit from as many constraints as possible on the orbital 
and eclipses parameters.
\end{itemize}

These data were used as input into an adaptative Markov Chain Monte Carlo (MCMC; see e.g. Tegmark 2004, Gregory 
2005, Ford 2006) algorithm. MCMC is a Bayesian inference method based on stochastic simulations that samples the 
posterior probability distribution of adjusted parameters for a given model. Our MCMC implementation uses the 
Metropolis-Hasting algorithm  (see e.g. Carlin \& Louis 2008) to perform this sampling. Our  model is based on a star and a transiting
 planet on a Keplerian orbit about their center of mass. More specifically, we used a classical Keplerian model for the RVs 
 obtained outside the transit in addition to a Rossiter-McLaughlin effect model (Gim\'enez 2006) for the RVs obtained 
 during transit. To model the eclipse photometry, we used the photometric eclipse model of Mandel \& Agol (2002) multiplied 
 by a systematic effect model different for each time-series (see Sec. 3.3). 
 
\subsection{Limb-darkening}

For both photometric transits, a quadratic limb-darkening law was assumed. For the FORS2 light curve, the quadratic coefficients  
$u_1$ and $u_2$ were kept fixed to 0.23 and 0.32, the values deduced from Claret'Õs tables (2000; 2004) for $T_{ef f}=5625$~K, 
$\log$~$g = 4.3$ and [Fe/H] = $0.0$ (B08). Considering the excellent quality of the 
CoRoT transit photometry, we allowed for it the quadratic coefficients $u_1$ and $u_2$ to float in our MCMC analysis, using 
as jump parameters\footnote{Jump parameters are the model parameters that are randomly perturbed at each step 
of the MCMC.} not these coefficients themselves but the combinations 
$c_1 = 2 \times u_1 + u_2$  and $c_2 = u_1 - 2 \times u_2$ to minimize the correlation of the obtained uncertainties (Holman et al. 2006). 
To obtain a limb-darkening solution consistent with theory, we decided to use a Bayesian penalty on $c_1$ and $c_2$
 based on theoretical values and errors for $u_1$ and $u_2$. The broad CoRoT band-pass does not 
correspond to any photometric filter, but its maximum of transmission is close to the V and R bands (Deleuil 
et al. 2008). We used the method described in Gillon et al. (2009b) to deduce from Claret'Õs tables (2000; 2004) the 
theoretical values for $u_1$ and $u_2$ and their errors $\sigma_{u_1}$ and $\sigma_{u_2}$ for the V and R filters and 
the spectroscopic parameters of CoRoT-2b reported in B08. For each 
coefficient, we took as initial value the mean of the values obtained for both filters. For the errors, we took the 
mean of the errors deduced for both filters and added it quadratically to the difference between both filters to take 
into account our ignorance of the effective wavelength of the photometry. We obtained this way $u_1 = 0.413 \pm 0.108$
and $u_2 = 0.293 \pm 0.038$ for our initial limb-darkening coefficients. Finally, the following Bayesian penalty was added to 
our merit function: \begin{equation}
BP_{\rm \textrm{ }  ld} = \sum_{i=1,2} \bigg(\frac{c_i - c'_i}{\sigma_{c'_i}} \bigg)^2\textrm{,}
\end{equation}
where $c'_i$ is the initial value deduced for the coefficient $c_i$ 
and $\sigma_{c'_i}$ is its error computed from $\sigma_{u_1}$ and $\sigma_{u_2}$ . 

For the spectroscopic transits, a quadratic limb-darkening law was also assumed. The values $u_1 = 0.465$ 
and $u_2 = 0.276$ were deduced from ClaretÕ's tables for the stellar parameters presented in B08 and for the V-filter, 
corresponding to the maximum of transmission of the HARPS and SOPHIE instruments. These values were 
kept fixed in the MCMC. 

\subsection{Modelled photometric systematic effects}

For each light curve, the eclipse model was multiplied by a trend model to take into account known low-frequency noise sources 
(instrumental and stellar). 

At 4.5 $\mu$m (InSb detector), the measured IRAC fluxes show a strong correlation with the position of the 
target star on the array. This effect is due to the inhomogeneous intra-pixel sensitivity of the detector and is 
now well-documented (see, e.g., Knutson et al. 2008 and references therein). Following Charbonneau et al. (2008), 
we modelled this effect with a quadratic function of the sub-pixel position of the PSF center: \begin{equation}
A(dx,dy) = a_1+a_2dx +a_3dx^2+a_4dy+a_5dy^2 +a_6dxdy \textrm{,}
\end{equation}
where $dx$ and $dy$ are the distance of the PSF center to the center of the pixel. Notice that we followed here 
D\'esert et al. (2009) and added the cross-term $a_6$ to the function $A(dx,dy)$. We measured the PSF center for 
CoRoT-2 in each image by fitting a Gaussian profile. Its $x$ position ranged from 192.89 to 193.05 during the run, 
while its $y$ position ranged from 240.12 to 240.39. 

At 8 $\mu$m (SiAs dectector), the intra-pixel sensitivity homogeneity is good, but another systematic 
affects the photometry. This effect is known as the `ramp' because it causes the gain to increase asymptotically over 
time for every pixel, with an amplitude depending on their illumination history (see e.g. Knutson et al. 2008 and 
references therein). Following Charbonneau et al. (2008) again,  we modelled this ramp as a quadratic function of $ln(dt)$:
\begin{equation} 
B(dt)  = b_1 + b_2 ln(dt) + b_3 (ln(dt))^2 \textrm{,} 
\end{equation}
where $dt$ is the elapsed time since 15 min before the start of the run. 

From the CoRoT photometry, the star CoRoT-2 is known to be variable at the 2-3\% level on a timescale 
of $\sim$ 4.5 days, corresponding to its rotational period (A08; Lanza et al. 2009). For the VLT and IRAC time-series, we 
modeled this low-frequency modulation by a time-dependent quadratic polynomial: 
\begin{equation}
C(dt) = c_1 + c_2 dt + c_3 dt^2 \textrm{,}
\end{equation}
where $dt$ is the elapsed time since 15 min before the start of the 
run. As the photometric modulation due to rotating spots 
is a wavelength-dependent effect, independent coefficients 
were fitted for the two IRAC time-series despite their covering of the same occultation.

At the end, the VLT/FORS2 trend model thus has three coefficients, the IRAC 4.5 $\mu$m 3 + 6 -1 = 8 coefficients and 
the IRAC 8 $\mu$m 3 + 3 -1 = 5 coefficients. All the used trend models are linear in their coefficients, so instead of 
considering these coefficients as jump parameters in the MCMC,  we choose to determine them
by linear least squares minimization at each step of the MCMC after division of the data 
by the eclipse model generated from the latest set of jump parameters (see Sec.~3.6). We used for this purpose the SVD method 
(Press et al. 1992), which has been found to be very robust. 

The CoRoT transit photometry is already corrected for known systematics and the stellar rotational variability. Nevertheless, we
preferred to do not assume it to be perfectly normalized and consider a flux normalisation factor $d_{norm}$ that was also 
determined via SVD at each step of the MCMC.

\subsection{Photometric correlated noise}

Taking into account the correlation of the noise is important to obtain reliable error bars on the fitted parameters (Pont et al. 2006). For this
purpose, we followed a procedure similar to the one described by Winn et al. (2008). For each light curve, the standard deviation of the residuals of the first chain  was determined for the best-fitting solution, without binning and with several time bins ranging from 10 to 30 minutes. For each binning, the following factor $\beta_{\rm \textrm{} red}$ was determined

\begin{equation}
\beta_{\rm \textrm{} red} =  \frac{\sigma_N}{\sigma_1} \sqrt{\frac{N(M-1)}{M}}\textrm{,}
\end{equation}
where $N$ is the mean number of points in each bin, $M$ is the number of bins, and $\sigma_1$ and $\sigma_N$ are respectively the standard deviation of the unbinned and binned residuals. The largest value obtained with the different binnings was used to multiply the error bars of the measurements. We obtained $\beta_{\rm \textrm{} red} =$ 2.2
for the VLT/FORS2 light curve, 1.25 for the IRAC/4.5$\mu$m curve, 1.14 for the IRAC/8$\mu$m curve and 1.25 for the CoRoT photometry. Thus, all the light curves show a significant level of correlated noise. The large $\beta_{\rm \textrm{} red}$ obtained for the FORS2 light curve can be attributed to the presence of spots on the stellar surface and the resulting increase of the correlation of the noise during the transit (see Fig. 3).

\subsection{Systemic RVs and jitter noise}

For each RV time-series, the systemic velocity was determined at each step of the MCMC from the residuals via SVD. 
Our code is able to account for more linear terms, i.e. for trends in the RV time-series, but it was not needed here.
For the RV data taken outside transit, we assumed a different systemic velocity for the SOPHIE and HARPS data to 
account for a possible difference of zero-point calibration between both instruments. Following B08, we added quadratically 
a jitter noise of 56 m~s$^{-1}$ to the errors to account for the stellar activity. For the spectroscopic transit data, we considered 
the same RV offset during the whole transit, so we did not add any jitter noise but only considered a 
different systemic velocity for both spectrographs. Our analysis of the residuals of the first MCMC chain showed
 us that the jitter noise of 56 m s$^{-1}$ assumed for the data taken outside transit was leading to a residual 
 $rms$ in good agreement with the mean error of the measurements. Still, we had to add an extra-noise of 13 
 m~s$^{-1}$ for the data taken during transit to obtain a similar agreement. The need for this extra-noise could 
 be explained by the inhomogenous surface of the spotted star and/or by the systematic errors brought by the measurement of
 the  RV via the fit of a Gaussian profile on the non symmetric cross correlation function of the spectrum (see Winn et al. 2005, 
Triaud et al. 2009).
 
 One could wonder why the correlation of the noise is
 not treated in a similar way for the spectroscopic and photometric eclipse time-series. The answer 
is that the time sampling of the RV time-series is much poorer than the one of the light curves and is similar to the timescale 
of ingress/egress. We can thus not estimate precisely the level of correlated noise at this frequency via the method described in Sec. 3.4. 
Still, the fact that the time sampling and the correlation timescale of the noise that we want to model  are similar makes the 
addition of the quadratic difference between the residual $rms$ and the  mean RV error a proper method to take into account this 
`red' noise.

\subsection{Jump parameters, priors and merit function}

The jump parameters in our MCMC simulation were: 
the planet/star area ratio $(R_p /R_s )^2$, the transit width 
(from first to last contact) $W$, the impact parameter 
$b' = a \cos{i}/R_\ast$, the two Lagrangian parameters $e \cos{\omega}$
and $e \sin{\omega}$ where $e$ is the orbital eccentricity and $\omega$ is 
the argument of periastron, and the $K_2$ parameter characterizing the amplitude
 of the orbital RV signal (see Gillon et al. 2009b). We assumed a uniform prior distribution
 for all these jump parameters.
 
 The products $V \sin{I} \cos{\beta}$ and $V \sin{I} \sin{\beta}$ were also jump parameters 
 in our MCMC, where $V \sin{I}$ is the projected stellar rotational velocity 
and $\beta$ is the spin-orbit angle (see Gim\'enez 2006). As we 
have an independent determination of $V \sin{I}$ from 
spectroscopy ($10.5\pm0.4$ km~s$^{-1}$, B08), we added the following 
Bayesian penatly to our merit function: \begin{equation}
BP_{\rm \textrm{ } V\sin{I}} = \frac{(V \sin{I} - V \sin{I}_{B08} )^2}{\sigma_{V\sin{I}_{B08}}^2} \textrm{,}
\end{equation} where $V\sin{I}_{B08}$ is 10.5 km~s$^{-1}$ 
and $\sigma_{V\sin{I}_{B08}}$ is 0.4 km~s$^{-1}$. 

A totally independent determination of the orbital period $P$ and time of 
minimum light $T_0$ was impossible because we folded the CoRoT transit 
photometry with the ephemeris presented in A08. This is why  we let these parameters 
vary under the control of the following Bayesian penalty:\begin{equation} 
BP_{\textrm{ }ephemeris} = \frac{(P - P_{A08} )^2}{\sigma_{P_{A08}}^2} +  \frac{(T - T_{A08} )^2}{\sigma_{T_{A08}}^2}\textrm{,}
\end{equation}
where $P_{A08}$ and $T_{A08}$ are the best-fitting values presented 
in A08 and $\sigma_{P_{A08}}$ and $\sigma_{T_{A08}}$ are their errors. In other words, we used a normal prior distribution
for these two jump parameters based on the CoRoT results reported in A08.

The merit function used in our analysis was the sum of 
the $\chi^2$ for each time-series and of the Bayesian 
penalties presented in Eq. 1, 6 and 7.

\subsection{Structure of the analysis}

Our analysis was similar to the one presented by Gillon et al. (2009b), consisting in four successive steps: 
\begin{enumerate}
\item First, we performed a single MCMC chain aiming to assess the level of correlated noise in the photometry and of jitter noise in the RVs and to update the
 measurement error bars accordingly. This chain was composed of $10^5$ steps, the first 20 \% of each chain being considered as its burn-in phase and discarded. 
\item Five new MCMC chains (10$^5$ steps each) were performed using the updated measurement error bars. The good convergence and mixing of these five chains 
was checked succesfully using the Gelman and Rubin (1992) statistic. The inferred value and error bars for each parameter were 
obtained from the marginalized posterior distribution. The goal of this second step was to provide us with an improved estimation of the stellar density $\rho_\ast = 1.31^{+0.04}_{-0.03}$ $\rho_\odot$. 
\item The deduced stellar density and the spectroscopic parameters presented in B08 were then used to determine the stellar mass $M_\ast$ and age $\tau_\ast$
 via a comparison with the stellar evolution models computed with the {\tt CL«ES} code (Scuflaire et al. 2008). We obtained a 
stellar mass $M_\ast = 0.96 \pm 0.08 M_\odot$ and a stellar age $\tau_\ast = 2.7^{+3.2}_{-2.7}$ Gyr. 
\item A new run of 20 MCMC chains was then performed. This step was identical to the second one, with the exception that at each step of the chains, the physical parameters $M_p$, $R_p$ and $R_\ast$ were computed from the relevant jump parameters and the stellar mass. For this latter, a value was randomly drawn at each step from the normal distribution $N(0.96,0.08^2) M_\odot$ derived in the previous step.
\end{enumerate}

\section{Results and discussion}

Table 1 shows the median values and 68.3\% probability interval for the jump and physical parameters given by our MCMC simulation, and compares them to the values presented in A08/B08. Notice that the planet/star flux ratios  reported in Table 1 are the deduced occultation depths corrected for the signal dilution due to the nearby star (see Sec.~2.1). Figure 4 shows the IRAC photometry corrected for the systematic and binned per five minutes, with the best-fitting eclipse model superimposed. The best-fitting models for the CoRoT and spectroscopic transits are presented in Fig. 5. 

\begin{table*}
\label{tab:params}
\begin{tabular}{lccccl}
\hline
Parameter  & This study & A08/B08 & Unit &\\ \noalign {\smallskip}
\hline \noalign {\smallskip}
$Jump$ $parameters$ & &  & \\ \noalign {\smallskip}
\hline \noalign {\smallskip}

Planet/star area ratio  $ (R_p/R_s)^2 $ & $0.02750 \pm 0.00012$ &   $0.02779 \pm 0.00020$ [A08]&  \\ \noalign {\smallskip}
$ b'=a\cos{i}/R_\ast $ & $ 0.223^{+ 0.018}_{- 0.020} $  &$0.253 \pm 0.012$ [A08] &  $R_*$  \\ \noalign {\smallskip}
Transit width  $W$ & $ 0.09446^{+0.00011}_{-0.00010}$  & -  &  days  \\ \noalign {\smallskip}
$e\cos{\omega}$ & $-0.00291^{+0.00063}_{-0.00061}$   & 0 (fixed) [A08 \& B08] & \\ \noalign {\smallskip}
$e\sin{\omega}$  & $0.0139^{+0.0079}_{-0.0084}$ &  0 (fixed) [A08 \& B08] & \\ \noalign {\smallskip}
$V\sin{I}\cos{\beta}$ & $10.79^{+0.33}_{-0.32}$ &   $11.76 \pm 0.51$ [B08]  & \\ \noalign {\smallskip}
$V\sin{I}\sin{\beta}$ & $-0.8^{+1.2}_{-1.1}$ &   $-1.48 \pm 0.93$ [B08] & \\ \noalign {\smallskip}
4.5 $\mu$m $dF_2$ & $0.00510^{+0.00041}_{-0.00040} $  &  - & \\ \noalign {\smallskip}
8 $\mu$m $dF_2$ & $0.0041 \pm 0.0011$  &  - & \\ \noalign {\smallskip}
RV $K_2$ & $725 \pm 22$ &  $678 \pm 17 $ [A08] &   \\ \noalign {\smallskip}
& & &  \\
$^*$Transit epoch  $ T_0$ & $ 2454237.53556^{+0.00020}_{-0.00021}$ &  2454237.53562 $\pm$ 0.00014 [A08] & HJD  \\ \noalign {\smallskip}
$^*$Orbital period  $ P$ & $ 1.7429935 \pm 0.0000010$ &  1.7429964  $\pm$ 0.0000017 [A08] &  days  \\ \noalign {\smallskip}
$^*$$c_1$ & $ 0.911 \pm 0.016$ & $0.88 \pm 0.07 $ [A08] &  \\ \noalign {\smallskip}
$^*$$c_2$  & $ -0.094 \pm 0.078$ &  $0.29 \pm 0.07$ [A08]  &  \\ \noalign {\smallskip}
\hline \noalign {\smallskip}
$Deduced$ $parameters$      &     &  &  \\ \noalign {\smallskip}
\hline \noalign {\smallskip}
$^*$$V\sin{I}$ & $10.87 \pm 0.32$ & $11.85 \pm 0.50$ [B08] & \kms \\ \noalign {\smallskip}
$\beta$ & $-4.0^{+6.1}_{-5.9}$ &$-7.2 \pm 4.5$ [B08]& degrees \\ \noalign {\smallskip}
RV $K$ & $ 603 \pm 18 $  & 563 $\pm$ 14 [A08] & \ms \\ \noalign {\smallskip}
$b_{transit}$ &   $ 0.221 ^{+ 0.017}_{- 0.019}$  &   $0.253 \pm 0.012$ [A08] &  $R_*$ \\ \noalign {\smallskip}
$b_{occultation}$ &   $ 0.226^{+ 0.018}_{- 0.020} $   &  $0.253 \pm 0.012$ [A08]  &  $R_*$ \\ \noalign {\smallskip}
$T_{occultation}$ & $ 2454238.40380 ^{+0.00071}_{-0.00068}$ &  2454238.40712 $\pm$ 0.00014 [A08] & HJD  \\ \noalign {\smallskip}
Orbital semi-major axis $ a $ & $ 0.02798^{+ 0.00076}_{- 0.00080} $  &  $0.0281 \pm 0.0009$ [A08]  & AU \\ \noalign {\smallskip}
Orbital inclination  $ i $ & $ 88.08^{+0.18 }_{-0.16} $  & 87.84 $\pm$ 0.1 [A08]  & degrees \\ \noalign {\smallskip}
Orbital eccentricity $ e $ & $ 0.0143^{+0.0077}_{-0.0076} $   &  0 [fixed, A08 \& B08]&  \\ \noalign {\smallskip}
Argument of periastron  $ \omega $ & $102^{+17}_{-5}$ &    - &  degrees  \\ \noalign {\smallskip}
Stellar mass $M_\ast $ & $ 0.96 \pm 0.08$ &   0.97 $\pm$ 0.06 [A08] & $M_\odot$  \\ \noalign {\smallskip}
Stellar radius  $ R_\ast $ & $ 0.906^{+ 0.026}_{- 0.027} $  & 0.902 $\pm$ 0.018 [A08] &  $R_\odot$ \\ \noalign {\smallskip}
Stellar density  $\rho_* $ & $1.288^{+ 0.035}_{- 0.033} $  & $1.327 \pm 0.006$ [A08] &  $\rho_\odot $\\ \noalign {\smallskip}
$u_1$ & $ 0.346^{+ 0.014}_{- 0.015}$ &  $0.41 \pm 0.03$ [A08] & \\ \noalign {\smallskip}
$u_2$ & $ 0.220 \pm 0.032$ & $0.06 \pm 0.03$ [A08] & \\ \noalign {\smallskip}
Planet mass  $ M_p $ & $ 3.47 \pm 0.22 $  &  3.31 $\pm$ 0.16  [A08] & $M_J$ \\ \noalign {\smallskip}
Planet radius  $ R_p $ & $ 1.466 ^{+ 0.042}_{- 0.044} $ & 1.465 $\pm$ 0.029 [A08]  & $R_J$ \\ \noalign {\smallskip}
Planet density  $ \rho_p $ & $1.105^{+0.060}_{+0.056}$ & $1.05 \pm 0.08$  [A08]&  $\rho_{J}$ \\ \noalign {\smallskip}
\hline\\ \noalign {\smallskip}
\end{tabular}
\caption{CoRoT-2 system parameters and 1-$\sigma$ error limits derived from our MCMC analysis. $^*$A Bayesian penalty was used for the parameters preceded by an asterisk. }
\end{table*}

\begin{figure}
\label{fig:d}
\centering                     
\includegraphics[width=8cm]{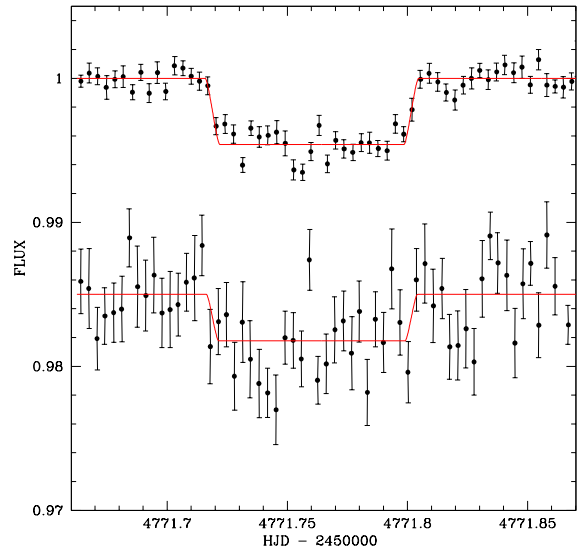}
\caption{IRAC 4.5 $\mu$m ($top$) and 8 $\mu$m ($bottom$) occultation 
photometry binned per five minutes and corrected for the systematics with the best-fitting 
occultation models superimposed. The bottom dataset is shifted for clarity. The dilution of the occultation due to the nearby fainter star is not corrected here. }
\end{figure}

\begin{figure}
\label{fig:e}
\centering                     
\includegraphics[width=8cm]{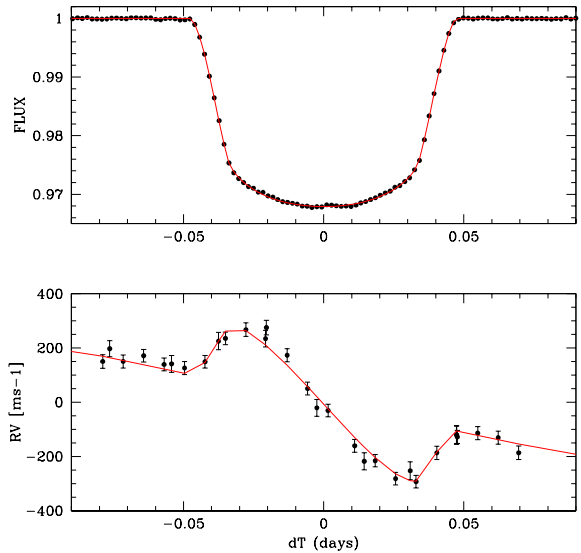}
\caption{$Top$: CoRoT transit photometry with the best-fitting transit model superimposed. $Bottom$: HARPS/SOPHIE transit RVs with the best-fitting RM model superimposed.  }
\end{figure}

\subsection{CoRoT-2b: a young, bloated and massive planet in a slightly eccentric and well-aligned orbit}

As shown in Table 1, our results for CoRoT-2b agree well with the results reported in A08 and B08.  Our error bars are in average larger than the ones reported in these previous works.  We consider our error bars as more reliable for the following reasons: (1) we did not assume a circular orbit, (2) we took into account the ephemeris errors for the folded CoRoT photometry, (3) we did not assume a perfect normalization for the CoRoT photometry, and (4) we took into account the correlated noise present in the light curves. To assess the impact of the VLT photometry on the final solution, we performed another MCMC integration without it, letting $P$ and $T_0$ free but under the control of a Bayesian prior based on A08 ephemeris. The resulting parameters and their error bars were all in good agreement with the values shown in Table 1, the main effect of the VLT light curve being to improve by a factor $\sim$ 2 the precision on the orbital period because it was obtained nearly one year after the end of the last CoRoT measurement, extending thus considerably the time baseline. 

We thus confirm the low density of the massive planet CoRoT-2b. Tight constraints 
on the age of the system could help understand this peculiarity. As shown in Sec.~3.7, stellar evolution modelling does not much constrain the age 
of CoRoT-2 ($\tau_\ast = 2.7^{+3.2}_{-2.7}$ Gyr). Still, we have for this system three different age indicators. First,
the presence of the Li I absorption line (B08) suggests a star still close to the ZAMS. We have obtained a new high-resolution high-SN spectrum of the star with the UVES spectrograph on the VLT (program 080.C-0661D, PI F.~Bouchy). The Li I line at 6707 \AA$\textrm{ }$ 
is  clearly detected in this spectrum (see Fig. 6). The Li I abundance measured from this line and MARCS atmospheric models (Gustafsson et al. 2008) is log n(Li) = +2.8. Basing on Sestito \& Randich (2005), this abundance suggests an age between 30 and 316 Myr. Secondly, the strong emission line core in the Ca II H and K lines observed in the series of HARPS spectra (B08) also indicates a young age. Using only the 14 HARPS spectra presenting a SN larger than 2 in the spectral region of the Ca II lines and the value 0.854 for the $B-V$ color of the star from the Exo-Dat database (Deleuil et al. 2009), we deduce a value of $-4.471 \pm 0.0629$ for the $\log(R^\prime_{HK})$ parameter. Following Wright et al. (2004), this activity leads to an age of 307 $\pm$ 150 Myr. Finally, the small rotation period of the star measured from the CoRoT light curve ($P_{rot}$ = 4.54 days, L09) and the $B-V$ color of the star inserted into the relationship presented in Barnes (2007) lead to an age of $76 \pm 7$ Myr for the system. As chromospheric activity is a magnetic phenomenon driven by rotation, it is clear that the $\log(R^\prime_{HK})$ is not an age indicator independent of the rotation period (e.g. Vaughan et al. 1981, Noyes et al. 1984). Still, we can conclude safely from both the large Li I abundance and  rotational velocity of the star that CoRoT-2 is a very young star, of a few hundreds of Myr at most. In this context, a part of the  `radius  excess' of the planet is explained. For instance, Fortney et al. (2007)  models of irradiated planets predict a radius of $\sim$ 1.3 $R_J$ for a planet  of 4 $M_J$ orbiting at 0.02 AU around a 300 Myr old solar-type star. 

Our occultation photometry imposes a strong constraint on the parameter $e \cos{\omega}$ and reveals that it is significantly smaller than zero. We can thus conclude that the orbit of CoRoT-2 is  slightly eccentric. Unfortunately, $e \sin{\omega}$ is much less constrained by our data, so the actual values of $e$ and $\omega$ are poorly known, as shown by their marginalized posterior distribution function (PDF, see Fig.~7). The PDF of $e$ itself is strongly not Gaussian: its 68.7\% and 99.9\% probability intervals are respectively $0.007 < e < 0.022$ and $0.001 < e  < 0.037$. To test the influence of the Bayesian penalty on $T_0$ and $P$ on the resulting PDF of $e \cos{\omega}$, we performed a new  analysis without using these penalties. We obtained $P=1.7429926\pm0.0000015$ and $e \cos{\omega}=-0.00258^{+0.00069}_{-0.00067}$. Thus the obtained period does not disagree significantly ($\sim$ 1.7 sigma) with the one obtained by A08 from the CoRoT photometry, while the offset of the occultation remains significative (3.7 sigma, $vs$ 4.6 sigma using the Bayesian penalties on $P$ and $T_0$). This offset could be due to a slight eccentricity of the orbit, but also to a dynamical interaction with another object in the system (see, e.g., Schneider 2004, Holman \& Murray 2005, Agol et al. 2005). Still, both the TTV analysis presented in Alonso et al. (2009a) and the agreement between our deduced period  and the one obtained by A08 argue against this last hypothesis. We conclude thus to a slight eccentricity of the planetary orbit. 

Our result for the spin-orbit projected angle $\beta$ agrees well with the one reported in B08, confirming a value close to zero for this parameter. To assess the influence of our Bayesian penalty on the parameter $V\sin{I}$, we performed another MCMC integration without this penalty. We obtained $V\sin{I}=11.6 \pm 0.5$ and $\beta = -5 \pm 7$ degrees, i.e. in agreement with the ones obtained with the Bayesian penalty, while beeing slightly less precise.

\begin{figure}
\label{fig:g}
\centering                     
\includegraphics[width=8cm]{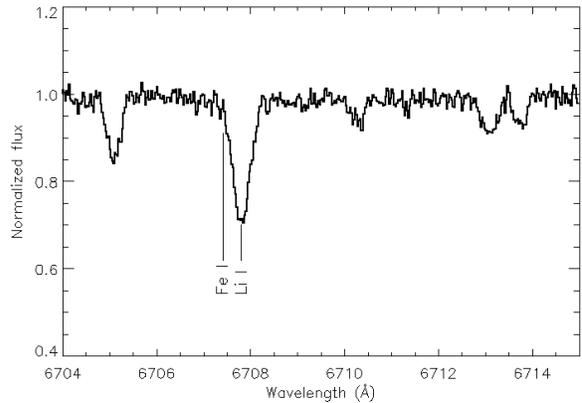}
\caption{Portion of the CoRoT-2 UVES spectrum showing the Li I line and the weak contaminating Fe I line.}
\end{figure}

\begin{figure}
\label{fig:g}
\centering                     
\includegraphics[width=6cm,angle=-90]{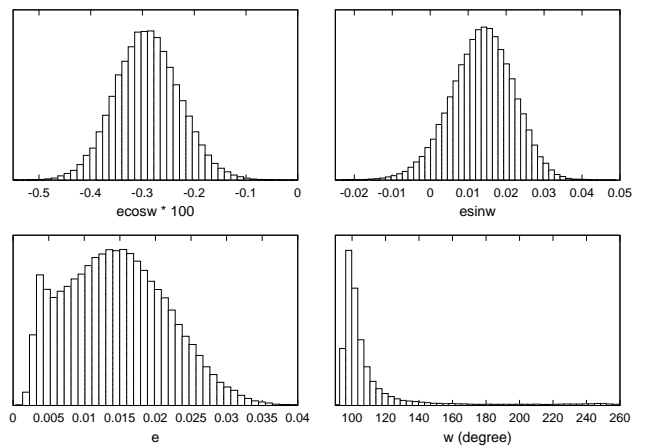}
\caption{Marginalized PDF obtained for (from top left to bottom right) $e\cos{\omega}$, $e\sin{\omega}$, $e$, and $\omega$. Notice how the PDF for $e$ and $\omega$ are non-Gaussian.}
\end{figure}

Interestingly, the $Spitzer$ flux estimator online tool\footnote{http://ssc.spitzer.caltech.edu/tools/starpet/} indicates that the differences in magnitude that we measured by deconvolution photometry between the nearby star and CoRoT-2 at 4.5 $\mu$m (+1.7) and 8 $\mu$m (+1.4) are consistent with a late-K or early-M type dwarf star located at the same distance ($\sim$ 200 pc) as CoRoT-2. As noticed in A08, this is also the case for the optical magnitude differences from the Exo-dat database and the 2MASS near-IR magnitudes. In case of gravitational bounding, the angular distance between both stars would correspond to a physical separation of $\sim$ 800 AU. It is thus desirable to assess this possible gravitational bounding by independent measurements (proper motion, radial velocity). In case of confirmation, CoRoT-2b would follow the tendency for massive planets to be found preferentially in multiple stellar systems (Eggenberger et al. 2004).

\subsection{Investigating the large radius of CoRoT-2b with coupled tidal-orbital evolution modelling }

CoRoT-2b is just one of many transiting planets with a radius larger than can be accomodated by standard thermal evolution models.  Given the relatively young age of the planet, compared to other known transiting planets, it is worthwhile to investigate the planet's radius evolution in some detail, as giant planets are expected to have larger radii at young ages.  We use the coupled giant planet tidal and thermal evolution model of Miller et al.~(2009) to calculate the planet's evolution and contraction.  As in Miller et al., the planet's structure is assumed to have three components: a 50\% rock 50\% ice core, a fully convective hydrogen-helium envelope with the equation of state of Saumon et al. (1995) and a non-grey atmosphere model
described by Fortney et al.~(2007).  The tidal orbital evolution is described by Jackson et al. (2008a, 2009).  This tidal evolution 
model assumes that the planet quickly reaches a spin-orbit sychronous state, that the only important source of tidal heating is due
to orbital circularization, and the model is second order in eccentricity.  

In order to determine if tidal heating can explain the large radius of CoRoT-2b, for a variety of tidal quality factors $Q_p'$, $Q_s'$, 
and core masses, a grid over initial semi-major axis and eccentricity are evolved forward in time.  
We search for instances in each of these evolution histories for which the semi-major axis, eccentricity, and radius are simultaneously within their error ranges.  
We choose to limit the age between 20 Myrs and 400 Myrs.  We find that in cases when the $Q_p'$ value is too high ($Q_p'$ of $10^6$ or $10^{6.5}$) 
there is not sufficient dissipation inside the planet to achieve the observed radius.
However, for the cases of $Q_s' = 10^5 - 10^6$ and  $Q_p'  \le 10^{5.5}$ all of the observed parameters can be explained as a transient event.  Evolution histories that closely agreed with the observed parameters are shown in Fig.~8 . 
The radius at optical wavelengths that the planet would be observed to have during the transit is shown in the upper left.  
The semi-major axis of the orbit is shown in the upper right.  The ratio of input tidal power to net radiated power is shown in the lower left.  The eccentricity is shown in the lower right.  (See Miller et al.~2009 for further details.)   
In each panel, the runs correspond to models with no core (black), 10 $M_\oplus$ core (red), 30 $M_\oplus$ core (blue).  A ``standard'' run without tidal effects with a 10 $M_\oplus$ core is in dotted cyan.  The model without tidal heating clearly cannot explain the planet's larger radius, even given the young system age. 
This analysis suggests that if the $Q_p'$ value is $10^{5.5}$ or smaller, then it is possible to explain this large radius as a transient event at the last stage of orbital circularization.  Under this scenario, the planet is spiralling inwards  at high speed to its final tidal disruption, and the fast rotation  of the star would not only be due to its young age but also to the high rate of angular momentum transfer from the planet's orbit. With such values for $Q_s'$ and $Q_p$,  the future lifetime of CoRoT-2b is 20 Myr at most, which is a short duration on an astronomical timescale but is still much larger than the remaining lifetime of the planet WASP-18b (Hellier et al. 2009) under similar assumptions.

In some planetary systems, an outer companion might continously drive the eccentricity of the inner planet offseting circularization by tides such that the eccentricy 
is found in a semi-equilibrium state, described by Mardling (2007).  Let us assume this scenario is occuring
and the planet's net radiated luminosity, $L_p$, at the surface is balanced by tidal heating inside, $P_t$.   Using Table 1 from Miller et al.~(2009)
\begin{eqnarray}
  L_p  & =  &7 \times 10^{28} \textrm{ ergs/sec} {}\nonumber\\
  & = & P_t =  \left\langle 4 \times 10^{27} \left(\frac{e}{0.01}\right)^2 \left(\frac{10^5}{Q_p'}\right) \right\rangle
\end{eqnarray}
and assuming that the observed eccentricity of 0.0142 is close to its equilibrium value, then this would imply that $Q_p' \sim 10^4$.  This is lower than the 
oft-quoted value for Jupiter between $10^5$ and $10^6$ (Goldreich \& Soter, 1966).  In summary, we find that a young age alone cannot explain the large radius of CoRoT-2b, but that plausible tidal heating evolutionary histories, with $Q_p' \sim 10^4 - 10^{5.5}$, can explain it.

\begin{figure}
\label{fig:tidalev}
\centering
\includegraphics[width=8cm]{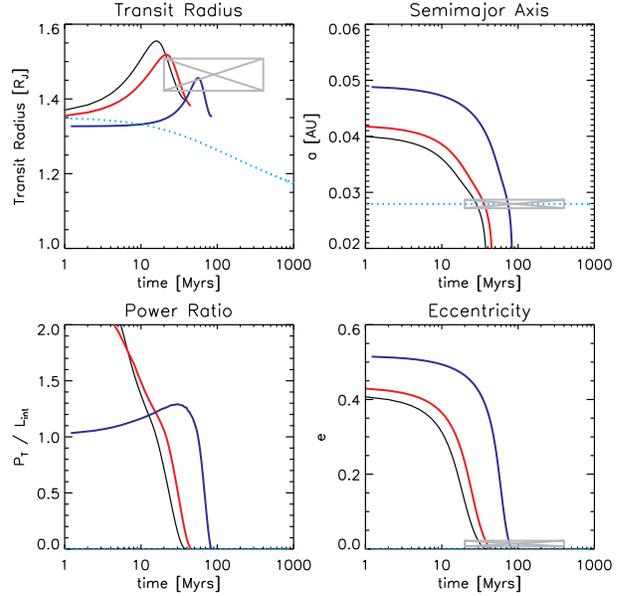}
\caption{Possible tidal evolution histories for CoRoT-2.  In these cases: $Q_p' = 10^{5.5}$ and $Q_s' = 10^5$.  For these curves we assume that the planet has no core (black), 10 $M_\oplus$ core (red) and 30 $M_\oplus$ core (blue).  The cyan run assumes that the planet has a 10 $M_\oplus$ core with no tidal evolution.  See text for discussion}
\end{figure}

\subsection{Atmospheric properties of the young planet CoRoT-2b}

The thermal emission of the planet is detected in both IRAC channels, as can be seen in Fig. 4 and 9. Unfortunately, the precision on the occultation depth at 8 $\mu$m is rather poor and thus brings a weak constraint on the planetary SED. CoRoT-2 is indeed a faint target for $Spitzer$ at 8 $\mu$m, the theoretical error (photon, read-out and background noise) per 10.4~s exposure being $\sim$ 0.84\%, while it is $\sim$ 0.29\% at 4.5 $\mu$m. The standard deviation of the residuals of our best-fitting solution are close to these values: 1.09\% (8 $\mu$m) and 0.32 \% (4.5 $\mu$m), i.e., respectively, 1.3 and 1.1 times larger than the theoretical noise budget. Assuming that the observed noise is purely white and taking into account the error on the flux normalization, we would expect an error of $\sim$ 0.06\% on the occultation depth at 8$\mu$m, while our MCMC analysis, which takes into account the low-frequency noises, leads to an error $\sim$ 1.8 times larger. We can thus conclude that the large level of  correlated noise (due to the ramp, the low-frequency stellar and background variability, the blend with the nearby fainter star, and other unknown effects)  has a significant effect on the final precision. This is also the case at 4.5 $\mu$m: for purely white noise, we would have expected a precision of 0.016\% on the occultation depth, while our actual error is $\sim$ 2.5 times larger. 

One of the many interesting questions that has arisen since the direct detection of hot-Jupiter atmospheres began (Charbonneau et al. 2005; Deming et al. 2005) is which, if any, of these planets has temperature inversions in their atmospheres (see Sec. 1).   So far, of the seven planets with published $Spitzer$ measurements in each of the four IRAC channels, all but HD 189733b and TrES-3b have been reported to exhibit temperature inversions (see references in Sec.~1). Figure 9 compares the IRAC colors\footnote{Here color is calculated
by taking the ratio of the planet fluxes in the IRAC channels and have not been
scaled by the flux ratios for Vega.} for these seven hot Jupiters.  
The lack of a
clear pattern in Fig.~10  illustrates the difficulty in using
$Spitzer$ photometry alone to identify an atmospheric inversion and highlights
the model dependencies of the inversions inferred so far from these data.  The
scatter in this diagram also demonstrates that an observationally based
classification scheme cannot yet be defined.

For CoRoT-2b, optical measurements are also available (Alonso et al. 2009b, Snellen et al. 2009b).
These latters and our two IRAC measurements are compared in Fig.~11 and Table 2 to three different models: \begin{itemize}
\item Model 1 ($m_1$) assuming an efficient heat distribution to the night-side of the planet.
\item Model 2 ($m_2$) assuming no heat distribution to the night-side and no temperature inversion.
\item Model 3 ($m_3$) assuming no heat distribution to the night-side and a deep TiO/VO-induced temperature inversion.
\end{itemize}

Table 2 shows also the $\chi^2$ obtained for each model. Given the $\chi^2$ 
of two models $m_a$ and $m_b$, we can compute their likelihood ratio:
\begin{equation} 
LR(m_a/m_b)  = e^{\frac{(\chi^2(m_a)-\chi^2(m_b))}{2}}
\end{equation}Comparing the model $m_1$ to the two others, we obtain $LR(m_1/m_2) = 3 \times 10^{-11}$ and  $LR(m_1/m_3) = 2 \times 10^{-9}$. 
An efficient heat distribution to the night-side of the planet is thus strongly disfavored by the data. Comparing now the models $m_2$ and $m_3$, we 
obtain $LR(m_2/m_3) = 74$. The data are thus in better agreement with the absence of a strong temperature inversion, but the obtained
likelihood ratio is not large enough to firmy conclude, and a greater wavelength coverage and a better precision in the observations are required  to remove the current ambiguity. Snellen et al. (2009b) concluded too that the CoRoT optical measurements favor an absence of temperature inversion. Nevertheless, we notice that their best-fitting model significantly under predicts the flux measured at 4.5 $\mu$m, this latter being in much better agreement with their models assuming an inversion. 

\begin{table*}
\label{tab:params}
\centering       
\begin{tabular}{lccccccl}
\hline\noalign {\smallskip}
  & CoRoT white & CoRoT red & IRAC 4.5 $\mu$m &  IRAC 8 $\mu$m  & $\chi^2$\\ \noalign {\smallskip}
\hline \noalign {\smallskip}
Measured & $0.0060 \pm 0.0020$$^\ast$ & $0.0102 \pm 0.0020$$^{\ast\ast}$ & $0.510 \pm 0.041$ & $0.41 \pm 0.11$&  \\ \noalign {\smallskip}
Model 1 & 0.0021&  0.0022 &  0.270  & 0.418  & 55.1 \\ \noalign {\smallskip}
Model 2 & 0.0062 &  0.0075 &  0.449  & 0.587  & 6.6 \\ \noalign {\smallskip}
Model 3 & 0.0068 & 0.0053 & 0.528 & 0.740 & 15.2 \\ \noalign {\smallskip}
\hline \noalign {\smallskip}
\end{tabular}
\caption{Comparison of the measured planet-to-star flux ratio and the values predicted by our three models. See text for details. $^\ast$from Alonso et al. (2009b). $^{\ast\ast}$from Snellen et al. (2009b). }
\end{table*}

\begin{figure}
\label{fig:g}
\centering                     
\includegraphics[width=5.5cm,angle=-90]{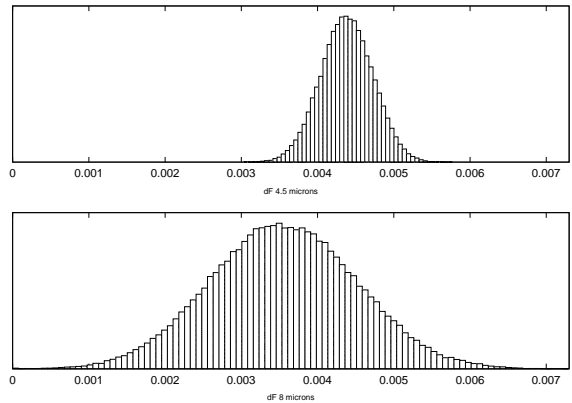}
\caption{Marginalized PDF for the IRAC occultation 
depth at 4.5 $\mu$m ($top$) and 8 $\mu$m ($bottom$).}
\end{figure}

\begin{figure}
\includegraphics[width=9cm]{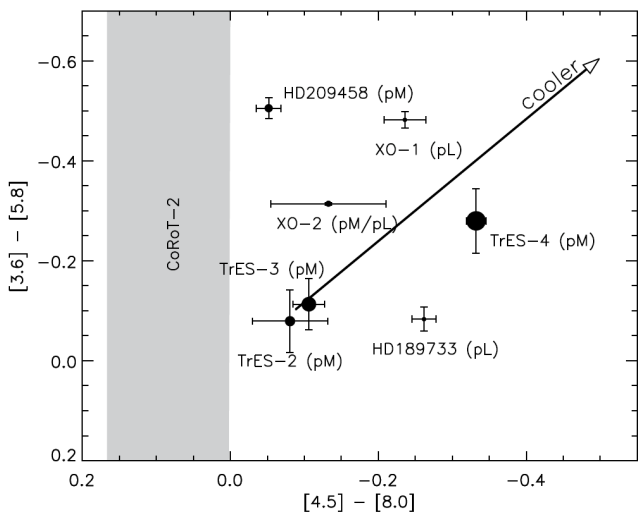}
\caption{IRAC color-color diagram for six hot Jupiters. Symbol sizes are scaled by the
level of incident starlight received by the planet.  The Fortney et al. (2008)
classification is also indicated.  The location of CoroT-2 falls in the shade region, 
indicating the 1-$\sigma$ uncertaities for the [4.5]-[8.0] color.
  The solid arrow indicates the trend for blackbody-emitting
planets.  See text for references.}
\label{colors}
\end{figure}

\begin{figure}
\includegraphics[width=9cm]{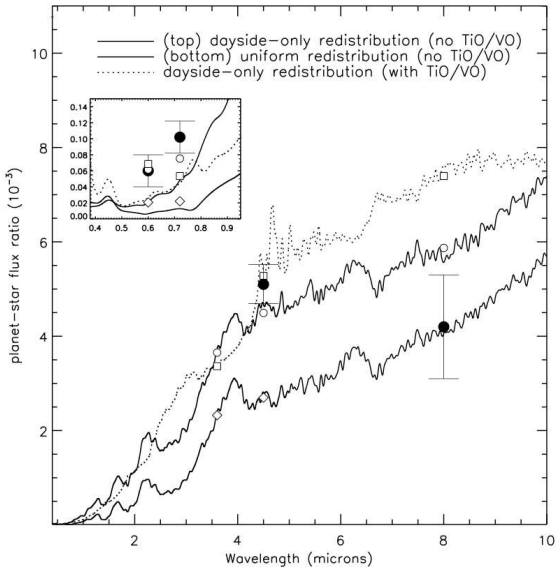}
\caption{Synthetic planet-star flux density ratios from three hot-Jupiter atmosphere
models (from Barman et al. 2005).  The top two curves are from models with incident stellar flux
constrained to the dayside, while the lower curve corresponds to a model with
unifom day-to-night redistribution of stellar flux.  The dotted line shows the
flux-ratio for a planet with a TiO/VO-induced temperature inversion.  Solid
points, with 1-$\sigma$ error-bars, are the $Spitzer$-IRAC and CoRoT photometry
(see inset).  Open symbols are the band-integrated model points.}
\label{fluxratio}
\end{figure}

\section{Conclusion}

Using $Spitzer$ and its IRAC camera, we measured an occultation of the young and massive planet CoRoT-2b
at 4.5 $\mu$m and 8 $\mu$m. In addition, we observed a partial transit of the planet with the Very Large Telescope 
and its FORS2 camera.

Our global analysis of CoRoT, $Spitzer$ and ground-based (FORS2 photometry + RVs) data confirms the low density of the planet ($\rho_p=1.105\pm0.060$~$\rho_{J}$ with $M_p =3.47\pm0.22$~$M_J$ and $R_p=1.466\pm0.044$~$R_J$). Constraining the system to be of at most a few hundreds of Myrs  old and the present orbit to be slightly eccentric ($e \cos{\omega} = -0.00291\pm0.00063$) and using coupled tidal-orbital evolution modelling, we find a self consistent thermal \& tidal evolution history that may explain the radius through a transient tidal circularization \& corresponding tidal heating inside the interior of the planet. Under this scenario, the planet will be tidally disrupted within 20 Myr at most.

The  occultation depths that we measured at 4.5 $\mu$m and 8 $\mu$m are, respectively, $0.510\pm0.042$ \% and $0.41\pm0.11$ \%. In addition to the optical measurements reported by Alonso et al. (2009b) and Snellen et al. (2009b), these values favor a poor heat distribution to the night side of the planet and the absence of thermal inversion, but measurements at other wavelengths are needed to confirm this latter point.

\begin{acknowledgements} 
This work is based in part on observations made with the $Spitzer$ $Space$ $Telescope$, which is operated by the Jet Propulsion Laboratory, California Institute of Technology under a contract with NASA. Support for this work was provided by NASA. The authors thank the ESO staff on the VLT  telescopes for their diligent and competent help during the observations. M. Gillon acknowledges support from the Belgian Science Policy Office in the form of a Return Grant. 

\end{acknowledgements} 

\bibliographystyle{aa}

\end{document}